

\long\def\UN#1{$\underline{{\vphantom{\hbox{#1}}}\smash{\hbox{#1}}}$}
\def\NP{\vfil\eject}
\def\NI{\noindent}

\magnification=\magstep
1
\overfullrule=0pt
\hfuzz=16pt
\voffset=0.0
true
in
\vsize=8.8
true
in
\baselineskip
20pt
\parskip
6pt
\hoffset=0.1
true
in
\hsize=6.3
true
in
\nopagenumbers
\pageno=1
\footline={\hfil
--
{\folio}
--
\hfil}

\hphantom{A}

\hphantom{A}

\centerline{\UN{\bf
Exact
Results
for
Diffusion-Limited
Reactions
with
Synchronous
Dynamics}}

\vskip
0.4in

\centerline{\bf
Vladimir
Privman}

\vskip
0.2in

\NI{\sl
Department
of
Physics,
Clarkson
University,
Potsdam,
New
York
13699--5820,
USA}

\vskip
0.4in

\centerline{\bf
ABSTRACT}

A
new
method
is
introduced
allowing
to
solve
exactly
the
reactions
A$+$A$\to$inert
and
A$+$A$\to$A
on
the
1D
lattice
with
synchronous
diffusional
dynamics
(simultaneous
hopping
of
all
particles).
Exact
connections
are
found
relating
densities
and
certain
correlation
properties
of
these
two
reactions
at
all
times.
Asymptotic
behavior
at
large
times
as
well
as
scaling
form
describing
the
regime
of
low
initial
density,
are
derived
explicitly.

\vfill

\NI

\NP

In
this
work
we
develop
a
new
method
of
obtaining
exact
results
for
certain
one-dimensional
reaction-diffusion
models.
Dynamics
of
the
1D
Ising
chain
was
first
solved
by
Glauber
[1].
This
famous
solution
has
lead
to
a
host
of
results
explored
over
the
last
three
decades.
However,
recently
there
has
been
a
resurgence
of
interest
in
the
solvable
1D
models
with
both
new
emphases
and
new
solution
methods,
developed
by
many
authors
[2-25].
The
new
emphases
have
been
on
those
models
which
have
no
equilibrium
states,
such
as
irreversible
reactions
[2-17],
cluster
coarsening
at
phase
separation
[17-23],
deposition
processes
[1,24],
models
of
self-organized
criticality
[25],
etc.
The
solution
methods
utilized
a
variety
of
different
approaches
both
for
continuous-time
asynchronous,
and
to
a
lesser
extent
for
synchronous
(discrete
time,
simultaneous
updating,
cellular-automaton-type)
dynamics
[17,26].

While
our
approach
can
be
applied
to
a
larger
class
of
models
[27],
the
present
work
is
devoted
to
the
diffusion-limited
(i.e.,
instantaneous
reaction
on
each
encounter)

coagulation,
A$+$A$\to$A,
and
annihilation
A$+$A$\to$inert,
of
particles
synchronously
hopping
on
the
1D
lattice.
The
models
will
be
defined
in
detail
later.
Besides
providing
interesting
examples
of
strongly
nonclassical
fluctuations,
1D
reactions
also
describe
certain
experimental
systems
[28]
with
the
reactants
being
typically
excitations
in
chain-like
structures.

Various
results
for
these
reactions
have
been
reported
in
the
literature
[2-17]
and
they
include
several
exact
solutions,
either
in
the
continuum
off-lattice
limit
[8-10,13]
or
on
the
lattice.
Our
results,
besides
providing
previously
unavailable
(on-lattice)
solutions
for
the
synchronous
case,
resolve
two
long-standing
theoretical
issues
in
this
field.
Firstly,
we
elucidate
and
prove
\UN{exactly}
the
equivalence
of
the
coagulation
and
annihilation
reactions
\UN{for
all
times
and
densities}.
This
equivalence
has
been
anticipated
and
explored
by
several
authors
[7-8,10,13-15]
based
on
the
asymptotic
large-time
results
and
on
certain
similarities
between
the
correlation
functions
of
both
reactions.
Here
we
derive
an
explicit
connection,
relations
(12)-(13)
below.

Secondly,
earlier
exact
on-lattice
calculations
for
A$+$A$\to$inert,
based
on
the
mapping
[4]
to
the
low-$T$
dynamics
of
Ising
interfaces
[4,11-12,17],
were
for
initial
conditions
corresponding
to
certain
subtle
correlations
in
the
particle
locations
at
time
$t=0$;
see
the
discussion
in
[12].
The
initially
uncorrelated
state
was
only
possible
for
the
(initial)
density
$\rho
=
{1
\over
2}$
(per
site).
As
a
result,
the
asymptotic
large-time
solution
was
not
universal.
Our
new
results
explicitly
apply
for
the
\UN{random}
initial
distribution
of
arbitrary
density
$\rho$
per
site.
The
large-time
behavior
of
both
the
coagulation
and
annihilation
reactions
is
found
to
be
universal
and
not
dependent
on
$\rho$.
The
onset
of
this
behavior
occurs
nonuniformly
for
small
$\rho$.
The
latter
regime
can
be
described
by
a
scaling
form
introduced
in
[29],
to
be
reviewed
later
on,
see
relations
(19)
and
(21)
below.
Our
exact
results
confirm
the
proposed
scaling
and
also
yield
explicit
form
of
the
appropriate
scaling
function.

We
consider
a
model
of
0
or
1
particles
at
lattice
sites
$i$
of
the
1D
lattice,
at
times
$t=0,1,2,\ldots$.
In
the
time
step
$t
\to
t+1$,
all
particles
hop
synchronously,
to
one
of
the
neighboring
sites
$i\pm
1$,
with
equal
probability
$1
\over
2$.
The
decision
which
way
to
hop
is
done
independently
for
each
particle.
However,
if
two
particles
end
up
at
the
same
site
they
instantaneously
``react''
according
to
the
annihilation
rule
(the
particle
number
is
reduced
from
2
to
0)
or
coagulation
rule
(the
number
is
reduced
from
2
to
1).
It
is
clear
that
this
dynamics
decouples
the
even-odd
and
odd-even
lattices.
Thus,
we
only
consider
particles
at
even
lattice
sites
$i=0,
\pm
2,
\pm
4,
\ldots$
at
even
times
$t=0,2,4,\ldots$,
and
at
odd
lattice
sites
$i=\pm
1,
\pm
3,
\ldots$
at
odd
times.
At
time
$t=0$
all
even
lattice
sites
are
randomly
occupied
(particle
number
1)
with
probability
$\rho
$
or
empty
(particle
number
0)
with
probability
$1-
\rho
$.

The
method
of
exact
solution
employed
here
utilizes
a
system
with
a
certain
conservation
property
which
allows
simplification
of
the
dynamical
behavior.
Thus,
rather
then
considering
particles
directly,
we
consider
integer
``charge''
variables
$q_i
(t)
=
0,
1,
2,
\ldots$
at
each
lattice
site
(with
the
even-odd
sublattice
convention
as
before).
The
charges
follow
the
dynamics
introduced
in
the
studies
of
self-organized
criticality
[25]:
they
randomly
hop
at
each
time
step,
similar
to
the
particles.
However,
if
two
charges
end
up
at
the
same
site
they
stick
together
and
move
as
a
single
charge
(sum
of
the
two
original
charges)
at
later
time
steps.
Such
dynamics
can
be
represented
by
the
rule
$$
q_i
(t+1)
=
a_{i-1}
(t)
q_{i-1}
(t)
+
\left[
1-
a_{i+1}
(t)
\right]
q_{i+1}
(t)
\;
,
\eqno(1)
$$
where
the
hopping-decision
variables
$a_j
(t)$
are
0
or
1
with
probability
$1
\over
2$.
Specifically,
$a_j
(t)=1$
indicates
that
the
charge
at
$j$
hopped
to
$j+1$
in
the
time
step
$t
\to
t+1$,
whereas
the
value
0
corresponds
to
stepping
to
$j-1$.

The
important
feature
of
such
dynamical
rules
[25]
is
that
due
to
charge
conservation
the
sum
of
charges
in
$r$
consecutive
sites,
$$
s_{r,j}
=
q_j
+
q_{j+2}
+
\ldots
+
q_{j+2(r-1)}
\;
,
\eqno(2)
$$
evolves
with
only
two
random
decisions
at
the
end-points
needed,
$$
s_{r,j}
(t+1)
=
a_{j-1}
(t)
q_{j-1}
(t)
+
q_{j+1}
(t)
+
\ldots
+
q_{j+2r-3}
(t)
+
\left[
1
-
a_{j+2r-1}
(t)
\right]
q_{j+2r-1}
(t)
\;
,
\eqno(3)
$$
where
the
random
variables
$a$
at
sites
$j-1$
and
$j+2r-1$
enter.
Relation
(3)
follows
directly
from
(1)
and
(2).

Consider
now
a
function
$F(s)$
defined
on
the
allowed
$s$
values
$0,1,2,\ldots$.
The
averages
$$f_r
(t)
=
\langle
F\left(s_{r,j}
(t)
\right)
\rangle
\;
,
\eqno(4)
$$
where
the
averaging
is
both
over
the
random
hopping
decisions
$a_i
(t)$
and
over
translationally
invariant
initial
conditions,
are
translationally
invariant,
i.e.,
they
do
not
depend
on
$j$.
By
using
relation
(3),
one
can
easily
verify
that
$$
f_r
(t+1)
=
{1
\over
4}
\left[
f_{r+1}
(t)
+
2
f_r
(t)
+
f_{r-1}
(t)
\right]
\;
,
\eqno(5)
$$
where
we
define
$f_0
(t)
=
F(0)$
to
have
(5)
apply
for
all
$r
\geq
1$.
Earlier
studies
used
this
formulation
with
the
choice
$F(s)
=
e^{i
\omega
s}$
and
a
further
complication
of
allowing
for
added
charge
``fed
in''
at
each
time
step,
by
including
inhomogeneous
term
in
(1)
with
values
drawn
from
a
fixed,
time-independent
distribution.
The
resulting
characteristic
functions
$f_r
(\omega,
t)$
can
be
used
to
study
the
self-similar
steady
state
and
dynamics
[27]
of
charge
systems
with
various
charge
variables,
---
real,
positive,
integer,
etc.,
---
in
the
framework
of
self-organized
criticality.

Our
approach
differs
in
that
we
employ
functions
$F(s)$
which
are
not
smooth.
Specifically,
let
us
define
the
\UN{empty
interval
indicator
function},
$$
F(s)=
\left\{
{{1,
\quad
s=0}\atop{0,
\quad
s>0}}
\right.
\;
\eqno(6)
$$
We
relate
the
particle
and
charge
systems
as
follows:
a
continuous
interval
of
$r$
sites
is
empty
of
particles
and
of
charge
simultaneously.
Thus,
we
associate
charge
$q>0$
at
a
given
site
with
particle
number
1,
and
charge
$q=0$
with
particle
number
0.
The
initial
state
of
0
or
1
particles
at
each
site
is
represented
by
$q_i(0)=0$
for
sites
$i$
empty
of
particles,
and
by
placing
an
arbitrary
initial
charge
$q_i(0)>0$
at
sites
occupied
by
particles.
Then
the
dynamics
of
charges
at
the
later
times
$t>0$
will
also
yield
the
correct
dynamics
of
\UN{coagulating}
particles
in
the
reaction
A$+$A$\to$A,
which
we
will
denote
by
the
subscript
A
for
brevity
in
what
follows.

The
initial
values
of
the
averages
are
$$
f_r
(0)
=
(1-
\rho
)^r
\;
,
\eqno(7)
$$
and
for
the
particle
density
per
site
we
get
$$
c_{_{\hbox{A}}}
(t)
=
1-
f_1
(t)
\;
,
\eqno(8)
$$
while
the
general
difference
$1-f_r(t)$
yields
the
probability
that
an
interval
of
$r$
consecutive
sites
is
empty.
Empty
interval
probabilities
have
been
considered
in
other
studies
of
the
coagulation
reaction,
e.g.,
[13].
Note
that
the
system
of
recursion
relations
(5)
for
the
coagulation
reaction
must
be
solved
with
the
initial
conditions
(7)
and
``boundary
condition''
$$
f_0
(t)
=
F(0)
=
1
\;
{}.
\eqno(9)
$$

Our
emphasis
here
will
be
on
the
density,
$c(t)$,
which
will
be
calculated
exactly.
However,
let
us
first
introduce
the
appropriate
formulation
for
the
annihilation
reaction
A$+$A$\to$inert,
which
will
be
denoted
by
the
subscript
{\O}
for
brevity.
The
idea
to
relate
the
aggregating
``charge''
system
to

hopping-particle
models
has
been
considered
in
[7,30].
Specifically,
Spouge
[7]
used
it
to
derive
various
exact
continuum-limit
expressions
for
both
types
of
reaction
(with
asynchronous
dynamics).
Besides
technical
differences,
the
implementation
here
is
more
powerful
than
earlier
variants.
Firstly,
synchronous
dynamics
and
the
associated
charge-conservation
property
(3)
allow
derivation
of
explicit,
discrete-time
exact
results
in
a
rather
straightforward
manner.
Secondly,
extensions
and
generalizations
are
more
easily
identified
[27].

For
the
annihilation
reaction,
we
consider
the
\UN{even-occupancy
indicator
function},
$$
F(s)=
\left\{
{{1,
\quad
s={\rm
even}}\atop{0,
\quad
s={\rm
odd}}}
\right.
\;
\eqno(10)
$$
Indeed,
the
mapping
to
the
particle
system
is
now
defined
by
associating
even
charges
with
empty
sites
and
odd
charges
with
sites
occupied
by
particles.
Initially,
each
empty
site
(particle
number
0)
is
assigned
an
arbitrary
\UN{even}
charge
value
$q(0)$,
whereas
every
occupied
site
(particle
number
1)
is
assigned
an
\UN{odd}
charge
value.
At
later
times
$t>0$
the
dynamics
of
charges
will
then
describe
the
annihilation
reaction
A$+$A$\to$inert.
Specifically,
$
c_{_{\hbox{\O}}}
(t)
=
1-
f_1
(t)
$,
while
the
difference
$1-f_r(t)$
denotes
the
probability
that
an
$r$-interval
has
an
even
number
of
particles
($0,2,4,\ldots,2[r/2]_{\rm
integer{\,}part}$)
in
it.

The
initial
conditions
for
the
annihilation
case
are
more
complicated
than
(7).
Indeed,
with
a
little
bit
of
combinatorics
one
can
check
that
the
correct
relation
is
$
f_r
(0)
=
\left[
1
+
(1-2\rho)^r
\right]
\big
/
2
$.
The
boundary
condition
(9)
is
unchanged.
This
suggests
consideration
of
the
modified
functions
$$
g_r
(t)
=
2f_r
(t)
-1
\;
{}.
\eqno(11)
$$
It
is
obvious
that
$g_r
(t)
$
satisfy,
for
{\O}
with
the
initial
density
$\rho_{_{\hbox{\O}}}$,
the
relations
(5),
(7)
and
(9)
exactly
identical
to
those
satisfied
by
the
original
averages
$f_r(t)$
for
the
reaction
A
provided
we
put
$$
\rho_{_{\hbox{A}}}
=
2
\rho_{_{\hbox{\O}}}
\;
{}.
\eqno(12)
$$

Thus
we
discover
that
for
the
particular
synchronous
dynamics
selected
for
the
annihilation
and
coagulation
reactions,
the
probabilities
of
finding
even
occupancy
for
the
former
are
related
to
the
probabilities
of
finding
empty
interval
for
the
latter,
with
twice
the
initial
density
in
coagulation
as
compared
to
annihilation.
The
precise
relations
can
be
quantified
via
the
identity
$
[g_r(t)]_{_{\hbox{{\O}}}}
=
[f_r(t)]_{_{\hbox{A}}}
$,
with
(12).
For
instance
for
the
densities
we
find
the
relation
$$
c_{_{\hbox{A}}}
(t)
=
2
c_{_{\hbox{\O}}}
(t)
\;
,
\eqno(13)
$$
which
for
this
particular
dynamics
is
\UN{exact
for
all
times}
$t=1,2,3,\ldots$
provided
the
random
initial
densities
satisfy
the
relation
(16)
at
$t=0$;
see
(12).
As
mentioned
earlier,
several
authors
have
explored
[7-8,10,13-15]
the
asymptotic
variants
of
this
result
for
other
dynamical
rules.

We
now
consider
the
coagulation
reaction
only
and
derive
exact
results
for
the
density
by
solving
the
recursions
(5),
with
(7)
and
(9),
by
the
generating
function
method.
Only
the
outline
of
the
actual
solution
steps
will
be
presented.
The
emphasis
will
be
on
discussion
of
results
for
the
density.
Thus,
we
consider
the
generating
functions
$\phi_r
(u)
=
\sum_{t=0}^\infty
f_r
(t)
u^t$
which
satisfy
the
relations
$\phi_0(u)=1/(1-u)$,
$\phi_r(0)=(1-\rho)^r$,
and
$$
u\left(
\phi_{r+1}+2\phi_r+\phi_{r-1}
\right)=
4\left[\phi_r-(1-\rho)^r
\right]
\;
{}.
\eqno(14)
$$

The
solution
of
(14)
is
obtained
by
first
eliminating
the
exponential-in-$r$
dependence
by
defining
$\phi_r=(1-\rho)^r
\psi_r$,
which
yields
an
autonomous
but
inhomogeneous
difference
equation
for

$\psi_r$.
However,
the
(constant)
inhomogeneous
term
is
then
removed
by
the
shift,
$\psi_r
=
\theta_r
+
\Theta$.
Finally,
the
resulting
second-order
difference
equation
for
$\theta_r$
is
solved
by
the
exponential
form,
$\theta_r=\Lambda^r
\theta_0$,
where
only
one
of
the
two
roots
of
the
characteristic
equation
gives
the
physically
acceptable
(i.e.,
regular
at
$u=0$)
$\Lambda$
value.
The
resulting
expression
is
$$
\phi_r=\left[
1
-
(1-u)^{1/2}
\right]^{2r}
u^{-r}
\left[
1/(1-u)
-
\Theta
\right]
+
(1-
\rho)^r
\Theta
\;
,
\eqno(15)
$$
$$
\Theta^{-1}=1-u(2-\rho)^2/[4(1-\rho)]
\;
{}.
\eqno(16)
$$

Let
us
now
analyze
the
generating
function
for
the
density,
$$1/(1-u)-\phi_1(u)=\sum_{t=0}^\infty
c(t)u^t=
2(1-u)^{-1/2}\left[1+(2-\rho)\rho^{-1}(1-u)^{1/2}
\right]^{-1}
\;
{}.
\eqno(17)$$
The
final
expression
was
obtained
from
(15)
with
(16)
and
required
a
rather
cumbersome
algebraic
manipulation.

Its
form
immediately
illustrates
that
the
large-time
density,
$$
c(t
\to
\infty)
=
2/\sqrt{\pi
t}
\;
,
\eqno(18)
$$
is
independent
of
$\rho$,
as
expected
by
the
universality
considerations
referred
to
in
the
opening
discussion;
the
residue
of
the
leading
singular
term
$\propto
(1-u)^{-1/2}$
does
not
depend
on
$\rho$.

We
also
note
that
the
onset
of
the
large-time
behavior
in
(17)
is
nonuniform
in
the
limit
$\rho
\to
0$.
Indeed,
it
has
been
argued
in
the
literature
[29]
that
the
universal
large-time
behavior
sets
in
after
the
initial
state
is
well
``mixed''
by
the
particle
diffusion.
The
(dimensional)
time
scale
for
such
a
mixing
is
of
order
[(dimensional)
density]$^{-2/{\rm
D}}/{\cal
D}$,
where
D
is
the
dimensionality
of
space
and
${\cal
D}$
is
the
diffusion
constant
of
the
particle
hopping.
Here
we
have
D$=1$
and,
in
our
dimensionless
units,
${\cal
D}=$O$(1)$.
Thus,
the
onset
of
the
large-time
behavior
in
the
limit
of
small
initial
densities
will
be
described
by
a
scaling
form
[29]
which
in
our
case
is
$$
c(t
\to
\infty,
\rho
\to
0)
=
t^{-1/2}
R(\rho^2
t)
\;
{}.
\eqno(19)
$$
The
prefactor
was
selected
to
have
the
correct
asymptotic
form
as
$t
\to
\infty$
for
fixed
$\rho>0$,
with
$R(x=\infty)=2/\sqrt{\pi}$,
while
for
short
times,
$t
\ll
{\rm
O}
\left(
\rho^{-2}
\right)
$,
the
reaction
is
ineffective:\
\
$R(x
\to
0)=\sqrt{x}$.

In
order
to
verify
these
expectations
we
need
an
explicit
expression
of
the
power
series
coefficients
of
(17).
These
can
be
obtained
in
terms
of
a
hypergeometric
function,
or
as
polynomials
in
$\rho$,
of
degree
$t+1$,
for
integer
$t$
values,
or
as
a
quadrature
[27].
We
favor
the
latter
expression,
$$
c(t)
=
{\rho
(2-\rho)^{2t+1}
(2t+1)
!!
\over
(1-\rho)^{t+1}
2^{3t+2}
t!
}
\int\limits_0^{4(1-\rho)
\over
(2-\rho)^2}
{z^t
dz
\over
(1-z)^{1/2}}
\;
,
\eqno(20)
$$
because
it
provides
an
explicit
analytic
continuation
to
all
real
$t
\geq
0$
values
(even
though
the
original
model
was
only
defined
for
integer
$t$).

Furthermore,
the
form
(20)
was
already
arranged
to
allow
a
relatively
simple
derivation
of
the
scaling
relation
(19).
We
find
$$
2R(x)=\sqrt{x\over
\pi}
\int\limits_0^\infty
{e^{-y/4}
dy
\over
(x+y)^{1/2}
}
\;
,
\eqno(21)
$$
which
satisfies
all
the
limiting
properties
as
expected.
Note
that
all
the
expressions
for
the
coagulation
reaction
are
meaningful
for
$0
\leq
\rho
\leq
1$.
However,
the
relation
to
the
annihilation
reaction,
(12),
(13),
etc.,
suggests
that
the
formulas
must
be
well
defined
for
all
$0
\leq
\rho
\leq
2$.
Indeed,
the
appropriate
results
for
the
annihilation
reaction
are
obtained
by
replacing
$\rho$
by
$2\rho$
in
all
the
``coagulation''
expressions
(and
of
course
adding
various
other
factors
such
as
the
coefficient
$1\over
2$
in
``translating''
the
density
expression).
Examination
of
(20)
reveals
that
the
powers
of
$1-\rho$
in
the
denominator
are
cancelled
by
the
appropriate
factors
from
the
integral
(note
$1-\rho$
in
the
upper
bound)
so
that
the
resulting
expressions
are
indeed
smooth
and
well
defined
near
$\rho=1$.

In
summary,
we
obtained
new
exact
results,
as
well
as
the
low-density
scaling
relation,
for
the
1D
coagulation
and
annihilation
reactions
with
synchronous
dynamics.
These
reactions
are
related
to
each
other
for
all
times.
The
large-time
expressions
are
universal.
We
remind
the
reader
that
the
notion
of
universality
refers
not
only
to
the
absence
of
dependence
of
results
such
as
(18)
on
the
initial
density
but
also
to
the
expectation
that
universal
results
apply
to
a
class
of
models
differing
only
in
microscopic
details
of
the
dynamics,
synchronous
or
asynchronous.
However,
for
latter
comparison
the
1D
densities
must
be
expressed
\UN{per
unit
length},
the
dimensionless
time
related
to
the
actual
physical
time
$\tau$,
etc.
For
instance
the
result
(18)
for
the
actual,
dimensional
density
of
the
coagulation
reaction
then
reduces
to
$(2\pi
{\cal
D}\tau
)^{-1/2}$,
etc.
Our
explicit
results
for
the
density
of
both
reactions
are
universal
when
thus
compared
with
other
calculations
available
in
the
literature
[2-17].

The
author
wishes
to
acknowledge
the
hospitality
and
financial
assistance
of
the
Institute
for
Theoretical
Physics
at
the
Technion,
Israel
Institute
of
Technology,
where
this
work
was
completed.

\NP\NI{REFERENCES}{\frenchspacing

\item{[1]}
R.J.
Glauber,
J.
Math.
Phys.
{\bf
4},
294
(1963).
Exact
results
for
certain
1D
adsorption
models
were
derived
by
E.R.
Cohen
and
H.
Reiss,
J.
Chem.
Phys.
{\bf
38},
680
(1963).

\item{[2]}
M.
Bramson
and
D.
Griffeath,
Ann.
Prob.
{\bf
8},
183
(1980).

\item{[3]}
D.C.
Torney
and
H.M.
McConnell,
J.
Phys.
Chem.
{\bf
87},
1941
(1983).

\item{[4]}
Z.
Racz,
Phys.
Rev.
Lett.
{\bf
55},
1707
(1985).

\item{[5]}
T.M.
Liggett,
{\sl
Interacting
Particle
Systems\/}
(Springer-Verlag,
New
York,
1985).

\item{[6]}
A.A.
Lushnikov,
Phys.
Lett.
A{\bf
120},
135
(1987).

\item{[7]}
J.L.
Spouge,
Phys.
Rev.
Lett.
{\bf
60},
871
(1988).

\item{[8]}
M.
Bramson
and
J.L.
Lebowitz,
Phys.
Rev.
Lett.
{\bf
61},
2397
(1988).

\item{[9]}
C.R.
Doering
and
D.
ben-Avraham,
Phys.
Rev.
A{\bf
38},
3035
(1988).

\item{[10]}
D.J.
Balding
and
N.J.B.
Green,
Phys.
Rev.
A{\bf
40},
4585
(1989).

\item{[11]}
J.G.
Amar
and
F.
Family,
Phys.
Rev.
A{\bf
41},
3258
(1990).

\item{[12]}
F.
Family
and
J.G.
Amar,
J.
Stat.
Phys.
{\bf
65},
1235
(1991).

\item{[13]}
D.
ben-Avraham,
M.A.
Burschka
and
C.R.
Doering,
J.
Stat.
Phys.
{\bf
60},
695
(1990).

\item{[14]}
J.-C.
Lin,
Phys.
Rev.
A{\bf
44},
6706
(1991).

\item{[15]}
J.-C.
Lin,
Phys.
Rev.
A{\bf
45},
3892
(1992).

\item{[16]}
V.
Privman,
Phys.
Rev.
A{\bf
46},
R6140
(1992).

\item{[17]}
V.
Privman,
J.
Stat.
Phys.
{\bf
69},
629
(1992).

\item{[18]}
A.J.
Bray,
J.
Phys.
A{\bf
23},
L67
(1990).

\item{[19]}
S.J.
Cornell,
K.
Kaski
and
R.B.
Stinchcombe,
Phys.
Rev.
B{\bf
44},
12263
(1991).

\item{[20]}
V.
Privman,
Phys.
Rev.
Lett.
{\bf
69},
3686
(1992).

\item{[21]}
P.L.
Krapivsky,
unpublished.

\item{[22]}
S.N.
Majumdar
and
C.
Sire,
unpublished.

\item{[23]}
J.-C.
Lin
and
P.L.
Taylor,
unpublished.

\item{[24]}
J.J.
Gonzalez,
P.C.
Hemmer

and
J.S.
H{\o}ye,
Chem.
Phys.
{\bf
3},
228
(1974).
Recent
results
have
been
reviewed
by
M.C.
Bartelt
and
V.
Privman,
Int.
J.
Mod.
Phys.
B{\bf
5},
2883
(1991),
and
J.W.
Evans,
Rev.
Mod.
Phys.,
in
print
(1993).

\item{[25]}
H.
Takayasu,
Phys.
Rev.
Lett.
{\bf
63},
2563
(1989),
S.N.
Majumdar
and
C.
Sire,
unpublished.

\item{[26]}
The
differences
between
the
asynchronous
and
synchronous
dynamics
are
usually
asymptotically
negligible
in
the
continuum
limit
typically
obtained
for
large
times
and
small
densities.
Some

interesting
aspects
of
different
dynamics
for
Ising
chains
near
$T=0$
have
been
elucidated
by
I.
Kanter,
Physica
D{\bf
45},
273
(1990).

\item{[27]}
V.
Privman,
unpublished.

\item{[28]}
Reviewed
by
R.
Kopelman,
C.S.
Li
and
Z.-Y.
Shi,
J.
Luminescence
{\bf
45},
40
(1990),
see
also
R.
Kroon,
H.
Fleurent
and
R.
Sprik,
Phys.
Rev.
E,
in
print
(1993).

\item{[29]}
K.
Kang,
P.
Meakin,
J.H.
Oh
and
S.
Redner,
J.
Phys.
A{\bf
17},
L665
(1984),
K.
Kang
and
S.
Redner,
Phys.
Rev.
A{\bf
32},
435
(1985).

\item{[30]}
B.
Derrida,
C.
Gordr\`
eche
and
I.
Yekutieli,
Phys.
Rev.
A{\bf
44},
6241
(1991).

}\bye